\documentstyle[aps,epsf]{revtex}
\newcommand{\re}{\mathop{\mathrm{Re}}}
\newcommand{\im}{\mathop{\mathrm{Im}}}
\newcommand{\Tr}{\mathop{\mathrm{Tr}}}
\newcommand{\diag}{\mathop{\mathrm{diag}}}

\newcommand{\Str}{\mathop{\mathrm{Str}}}

\begin{document}
\author{
Yan V. Fyodorov$\S\P$\thanks{
On leave from Petersburg Nuclear Physics Institute, Gatchina
188350, Russia}}
\address{
$\S$Fachbereich Physik, Universit\"at-GH Essen,
D-45117 Essen, Germany
        }
\address{
$\P$ MPI f\"{u}r Komplexer Systeme, Dresden 01187, Germany }

\title{ Spectral Properties of Random Reactance Networks
and Random Matrix Pencils}

\date{June 6, 1999}
\maketitle
\begin{abstract}
Our goal is to study statistical properies of 
 "dielectric resonances" which are poles of 
conductance of a large random $LC$ network. 
Such poles are a particular example of eigenvalues $\lambda_n$ 
of {\it matrix pencils} ${\bf H}-\lambda {\bf W}$,
with ${\bf W}$ being a positive definite matrix and ${\bf H}$ a random 
real symmetric one.
We first consider spectra of the
 matrix pencils with independent, identically distributed
 entries of ${\bf H}$. Then we concentrate on
an infinite-range ("full-connectivity") version of random $LC$ network. 
In all cases we calculate the mean eigenvalue density and
the two-point correlation function in the framework of Efetov's
supersymmetry approach. Fluctuations in spectra turn out to be
the same as those provided by the Wigner-Dyson theory of usual random matrices. 

\end{abstract}

\section{Introduction}

Large random reactance networks (that is networks made of 
random mixture of capacitances $C$ and inductances $L$) possess a peculiar
property noticed first by Dykhne \cite{Dykhne}: they have
a finite real conductance and thus can disperse an electric power.
Explanation of this somewhat paradoxical property is simple,
however. Indeed, if such a network is large enough
there always exist circuits of "resonance type", with 
(purely imaginary) conductance 
showing poles at some frequencies. Then the real part of the
conductance as a function of frequency 
consists of a set of $\delta-$like peaks at those resonance frequencies.
 When the volume of the network grows to infinity, 
this set becomes more and more dense.
 Then, adding an arbitrary small (infinitesemal, but fixed) active part to all 
inductances ( one may think, e.g.
of the inductance $L$ on each bond being in series with a weak resistance R)
results in a finite active resistance of the network. 

A random mixture of two active conductances
is very well studied in correspondence with 
the bond percolation problem, see e.g.\cite{Luck1}.
At the same time, the random reactance networks are 
relatively less studied.

It is necessary to mention that the random $LC$ (more generally, $RL-C$)
networks emerge in various physical contexts. As was shown 
long ago by Shender\cite{Zhenya}
the conductance of the random $LC$ network turns out
to be intimately related to properties of collective excitations
in spin glasses. The mapping between the two problems is 
possible by an analogy noticed first by Kirkpatrick \cite{kirk} 
between the Kirchhoff's law and the equation
of motion for the spin operators.

More recently, $RL-C$ networks
were claimed to be an adequate model for describing the 
optical absorption in disordered metal films, see \cite{BBS} 
and references therein, which showed some unusual features. 
This fact motivated Luck and collaborators 
for a series of insightful numerical investigations of
two-dimensional $RL-C$ arrays \cite{Luck2,Luck3}.

Since it is the conductance poles (resonances) 
of the $LC$ networks which dominate properties of the weakly dissipative
$RL-C$ networks, it is natural to try to understand
their properties in a greater detail. 
To access
 those poles it is convenient, following \cite{Luck3,Straley}
to start with the Kirchhoff
equations for the electric potential at vertices $i$
of a network:
\begin{equation}\label{Khhof}
\sum_j\sigma_{ij}(V_i-V_j)=0
\end{equation}
where the summation goes over all vertices $j$ which are neighbours
of a given $i$. The equations Eqs.(\ref{Khhof}) should be 
complimented with boundary conditions at electrodes. 
For a two-terminal geometry one assumes $V_i=0$ (resp. $V_i=V$)
for the vertices $i$ belonging to the left (resp. right) terminal. 

It is useful to introduce the 
(positive semidefinite) Laplace operator ${\bf D}$ on the network
by 
\begin{equation}\label{Lap}
({\bf D}V)_i=\sum_j(V_j-V_i)
\end{equation}
with a convention $V_i=0$ on both electrodes.

In a random $LC$ network each conductance at frequency
$f=\omega/2\pi$ is equal to either $\sigma_0=iC\omega$
or $\sigma_1=(iL\omega)^{-1}$, with a specified probabilities
(in what follows we concentrate on the case of equal probability
for finding $L$ and $C$ bonds in the network). Then the Laplace operator
can be written as a sum $\bf{D}_C+{\bf D}_L$ of its components
on the $L-$ and $C-$ bond sets, respectively. It is easy to show
that the poles of the conductance occur
at frequencies $f$ given by the roots of the equation \cite{Luck3,Straley}:
\begin{equation}\label{pendef}
\det{\left({\bf D}_L-\lambda{\bf D}\right)}=0
\end{equation}
where we introduce the ratio 
\begin{equation}\label{lambda}
\lambda=\frac{\sigma_0}{\sigma_0-\sigma_1}=
\frac{(\omega/\omega_0)^2}{1+(\omega/\omega_0)^2}\quad;\quad 
0\le \lambda\le 1
\end{equation}
with $\omega_0=1/(LC)^{1/2}$ being a characteristic 
resonant frequency of the LC network.

The simplest, but very informative way
 to understand properties of the random binary mixtures 
{\it on average} is to write an equation for the mean 
conductance $\Sigma$ using an effective-medium approximation (EMA).
For a network with
a coordination number $z$ and equal concentrations of $\sigma_0-$ and 
$\sigma_1-$bonds it reads as follows\cite{kirk}:
\begin{equation}\label{ema}
\frac{\Sigma-\sigma_0}{2\sigma_0+(z-2)\Sigma}+
\frac{\Sigma-\sigma_1}{2\sigma_1+(z-2)\Sigma}=0
\end{equation}
Analysing the solution of this quadratic equation as a function
of $\lambda$ in the interval $0\le \lambda\le 1$, 
one finds that generically for $z>4$ the real part of $\Sigma(\lambda)$
is non-zero only inside some interval 
$\lambda_{min}(z)\le \lambda \le \lambda_{max}(z)$.
According to the discussion above it means that a mean density of
resonances is finite only inside that interval. 

Useful and simple as it is, EMA  
suffers from an essential drawback: it systematically neglects
fluctuations, whereas taking the fluctuations into account
could be important. 
For example, as was noted in \cite{Zhenya,Luck3} the existence of sharp
edges $\lambda_{min,max}$ was an artifact of the EMA approximation.
In fact, the density of the resonances is never exactly zero inside
the whole interval $\lambda\in [0,1]$ due to the so called Lifshitz
tails, which is a purely fluctuation phenomenon. 
Apart from that, it seems hard to construct EMA as a systematic
approximation with respect to some small parameter, though there 
are indications that it becomes progressively exact for higher spatial
dimensions $d$ \cite{LS} and can be an extremely well-working one
for many realistic systems \cite{kirk}. 
Therefore, it is not clear how to correct it systematically
or to include fluctuation effects into account.

These unsatisfactory features should be contrasted 
with the status of the mean-field approximation
in the theory of magnetism, with which EMA shares its main
ingredients. As is well known, the mean-field 
equations become exact for the model with the infinite range of spin
interaction. Even for strongly disordered spin systems like spin
glasses the latter model provides an adequate basis for constructing
a mean-field theory with many non-trivial properties\cite{sglas}. 

It is natural to try to consider a model of similar type
 also for the disordered reactance networks. This just amounts to 
considering the Kirchhoff equations Eqs.(\ref{Khhof}) on a disordered 
{\it full-connectivity} graph of $N$ vertices (nodes)
connected by $N(N-1)/2$ edges (bonds), each independently
taking a value $\sigma_0$ or $\sigma_1$. 

Actually, we find it convenient to
 consider a graph with $N+2$ nodes, 
among them two "terminal" nodes 
(labeled $A$ and $B$, respectively) are singled out by being attached to
external voltage, so that the potential at $A$ is equal to $V$,
whereas the potential at $B$ is kept zero.
The rest of $N$ "internal" nodes are labeled by
index $i=1,...,N$ and the corresponding (induced) potentials are
denoted with $V_i$. The nodes are connected in a full-connectivity graph of
$(N+2)(N+1)/2-1$ bonds, with a single bond being excluded by obvious
reasons: that connecting terminals $A$ and $B$ directly, see Fig.1. 
%%%%%
\begin{figure}
\centerline{\epsfxsize=8cm
\epsfbox{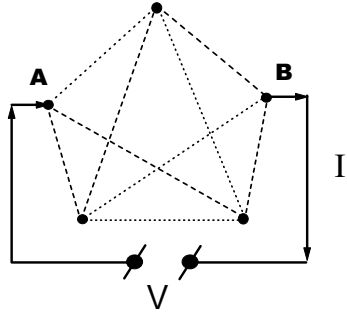}}
\caption{ Schematic view of the full-connectivity graph with three internal nodes. One direct bond connecting the terminal nodes $A$ and $B$ 
is excluded.
 \label{fig:tdelay}  } \end{figure}

%%%%%%
We then attribute
conductances $\sigma_{\mu\nu}=\sigma_{\nu\mu},
\,\, \mu,\nu=A,B,1,...,N$ to each bond.
It is convenient to introduce three $N-$ component vectors:
$ \underline{v}=(V_1,...,V_N)^T,\,\,
\underline{\sigma}_A=(\sigma_{A1},...,\sigma_{AN})^T,\,\, 
\underline{\sigma}_B=(\sigma_{B1},...,\sigma_{BN})^T$ (we used here $T$ to 
indicate the transposition)
and $N\times N$ matrix ${\bf \Sigma}$ with the following structure:
\[ 
{\bf\Sigma}=\left(\begin{array}{cccc} \sigma_{A1}+\sigma_{B1}+\sum_{k\ne 1}^N
\sigma_{1k}& -\sigma_{12}& ... & -\sigma_{1N}\\
-\sigma_{21}&\sigma_{A2}+\sigma_{B2}+\sum_{k\ne 2}^N
\sigma_{2k}& ... & -\sigma_{2N}\\
... & ... & ... &... \\
-\sigma_{N1}&-\sigma_{N2}&
 ... &\sigma_{AN}+\sigma_{BN}+\sum_{k\ne N}
\sigma_{Nk}\end{array}\right)
\]
In these notations $N+1$ Kirchhof's equations Eq.(\ref{Khhof}) 
($N$ for the
internal nodes $i$ and one for the node $B$) can be written, respectively:
\begin{equation}\label{kin}
{\bf \Sigma}\underline{v}=V\underline{\sigma}_A\quad,\quad 
\underline{\sigma}^T_B \underline{v}=I
\end{equation}
where $I$ stands for the total current flowing outwards
through the terminal node $B$. This system of equations can be readily solved
yielding the expression for the network conductance:
\begin{equation}\label{Y}
Y_{AB}=\frac{I}{V}=\underline{\sigma}^T_B{\bf
\Sigma}^{-1}\underline{\sigma}_A 
\end{equation}

If we now deal with a binary networks when each $\sigma_{ij}$
is $\sigma_0$ with a probability $p$ and $\sigma_1$ with the probability
$1-p$ it is convenient to introduce the "symmetric" 
variables $h_{ij}$ such that
$h_{ij}=-1$ if $\sigma_{ij}=\sigma_0$ and $h_{ij}=1$ if
$\sigma_{ij}=\sigma_1$, so that $\sigma_{ij}=\frac{1}{2}
\left([\sigma_0+\sigma_1]+[\sigma_1-\sigma_0]h_{ij}\right)$.
In terms of these variables the network conductance Eq.(\ref{Y})
can be written as:
\begin{equation}\label{YY}
y_{ab}=Y_{AB}/\sigma_0=-\frac{1}{2\lambda}
\left[\underline{h}_B^T-\tilde{\lambda}\underline{e}^T\right]
\frac{1}{{\bf H}-\tilde{\lambda}{\bf W}}
\left[\underline{h}_A-\tilde{\lambda}\underline{e}\right]
\end{equation}
where $\tilde{\lambda}=2\lambda-1,\quad \underline{e}^T=(1,1,...,1)
,\quad \underline{h}_B^T=(h_{B1},...,h_{BN})$ and 
$N\times N$ matrices ${\bf W, H}$ have the following elements:
\begin{equation}\label{defin}
W_{ij}=(N+2)\delta_{ij}-1\quad,\quad
H_{ij}=\delta_{ij}\left(h_{Ai}+h_{iB}+\sum_{k\ne i} 
h_{ik}\right)-(1-\delta_{ij})h_{ij}
\end{equation}
With these expressions in hand we see that resonances of our network are
determined by values of $\tilde{\lambda}$ satisfying the condition:
\begin{equation}\label{reson}
\det\left({\bf H}-\tilde{\lambda}{\bf W}\right)=0
\end{equation}

Of course, it is evident that the full-connectivity
construction would be completely inadequate one for describing
the percolation problem which dominates properties of binary
mixtures of two real conductances. 
We are, however, interested in studying the resonances of disordered
LC networks, and the existence of such resonances is in no way 
precluded by "all-to-all" geometry. We will find, that such a model 
turns out to be, in essence, 
exactly soluable in the limit $N\gg 1$. 
We start with deriving the mean density of resonances 
which turns out to be a function of the scaled variable 
$r=\tilde{\lambda}N^{1/2}$. One can envisage such a scaling already from
the EMA result Eq.(\ref{ema}), which for $z\gg 1$ gives 
$|\lambda_{min,max}-1/2|\propto
\tilde{\lambda}_{min,max}\sim z^{-1/2}$. At the same time, 
in contrast to EMA the support of the spectrum in 
the infinite-range model  does
not have artificial sharp edges $\tilde{\lambda}_{min,max}$, 
but rather the mean spectral density $\rho(r)$ smoothly decays to 
zero as long as $r\to \infty$ as $\rho(r)\propto e^{-r^2/2}$.

More interesting is the fact, that the infinite-range model
opens a possibility to study fluctuations of various quantities.
In the present paper we concentrate on spectral fluctuations and,
correspondingly, study the two-point correlation function of the resonance
densities. The latter turns out to be essentially the same as 
given by the famous Wigner-Dyson theory of random matrix spectra.
This fact favourably agrees with overall
numerical results \cite{Luck3} 
for the resonance spectra of two-dimensional disordered $LC$ networks. 
An origin of relatively small, but noticable 
deviations from the Wigner-Dyson statistics detected in \cite{Luck3}
remains unclear for us at the moment and could be related to $2D$
features of the  networks studied there which are not captured 
adequately by our infinite-ranged model. This issue
deserves further investigations. 
  
Actually, finding the set of $\lambda$'s satisfing Eq.(\ref{pendef})
or Eq.(\ref{reson})
is an example of the generalized eigenvalue problem. The combination
of matrices ${\bf D}_L-\lambda{\bf D}$ or ${\bf H}-\tilde{\lambda}{\bf W}$
in this respect is known 
in the mathematical literature as a {\it pencil of matrices}\cite{gant}
or just the matrix pencil. 
Theory of the matrix pencils has many important applications such as e.g.
 vibration and bifurcation analysis in complicated structures
\cite{pencils} and game theory\cite{games}. 

At the same time, it seems that the present
 knowledge on statistical properties
of generalized eigenvalues of the pencils of random matrices is rather
scarce. The paper \cite{EKS}
considers the mean number and the density of real eigenvalues
for a pencil ${\bf H}-\lambda {\bf W}$, with both ${\bf H}$ and 
${\bf W}$ being matrices with all independent real entries 
and no symmetry conditions imposed. At the same time,
our original physical problem motivates us to be
interested in the pencils formed by real {\it symmetric} 
matrices ${\bf H}, {\bf W}$, with 
${\bf W}$ being positive definite. It is clear
that ${\bf W}>0$ ensures all the eigenvalues of 
the matrix pencils to be real (such pencils are sometimes called in the
literature  the {\it regular} ones \cite{gant}).
Indeed, in that case the generalized eigenproblem: ${\bf H} \underline{x}=
\lambda {\bf W} \underline{x}$ is equivalent to a usual one:
$ {\bf W}^{-1/2}{\bf H}{\bf W}^{-1/2}
\underline{y}=\lambda\underline{y}\quad,
\quad \underline{y}={\bf W}^{1/2} \underline{x}$, with $\tilde{\bf H}=
{\bf W}^{-1/2}{\bf H}{\bf W}^{-1/2}$ being a real symmetric matrix.

The mean eigenvalue density for matrices of similar types
(when both ${\bf W}$ and ${\bf H}$ are random) was studied in some 
generality starting from the work by Marchenko and Pastur, see \cite{MP}.
In fact, the mean eigenvalue density can be found for the ensemble
 $\tilde{\bf H}$, with
${\bf H}$ being a real symmetric matrix with independent, identically
distributed (i.i.d.) entries using a generalized version of
results by Pastur\cite{Pas}.
 We are however not aware of any systematic study of spectral
correlations of regular pencils of the random matrices.

This should be contrasted with a very intensive research
on eigenvalues of the random matrices (which is a particular case
of ${\bf W}\equiv{\bf 1}$)
performed in recent years in the domain of 
theoretical and mathematical physics\cite{Guhr}.
Below we summarize the most important facts known from these
studies.
 
As is well established \cite{Pas1,univ1,HW,MF,kkp},
the statistical properties of real eigenvalues
$X_{i}$ of large $N\times N$ self-adjoint random
matrices ${\bf H}$ are to large extent {\it universal}, i.e.
independent of the details of the distributions ${\cal P}({\bf H})$
of their entries.

It is important to mention the existence, in general, of two
different characteristic scales in the random matrix spectra: the {\it
global} one and the {\it local} one. The global scale is that on which
the eigenvalue density, defined as 
$\rho(X)=\frac{1}{N}\mbox{Tr}\delta(X-{\bf H})$,
 changes appreciably
with its argument $X$ when averaged over ${\cal P}({\bf H})$. For matrices  
whose spectrum has a finite support
in the interval $X\in [A,B]$ the global scale is just the length of
this interval.

In contrast, the local
scale is that determined by the typical
separation $\Delta=\langle X_i-X_{i-1}\rangle$
between neighbouring eigenvalues situated
 around a point $X$, with the brackets standing for the statistical averaging.
It is given therefore by $\Delta=(\langle N\rho(X)\rangle)^{-1}$.
If we are interested in those values of $X$ that are sufficiently far
from the edges of the spectra the global scale is , roughly speaking,
by a factor of $N$ larger than the local one.
In other words, the mean density $\langle\rho(X)\rangle$
can be considered as a constant one on the scale $\Delta$.

The degree of universality is essentially
 dependent on the chosen scale.

As to the global scale universality,
one can mention first of all that  for the matrices with i.i.d. entries
under quite general conditions (see, e.g., \cite{semi} 
and references therein) the mean density
 is given in the limit $N\to \infty$ by the so-called "Wigner semicircle law":
\begin{equation}\label{semi}
\langle\rho(X)\rangle=\frac{1}{2\pi a^2 }
\sqrt{4\,a^2-X^2}=\frac{1}{N\Delta}\end{equation}
 In this expression the parameter $a$ just sets the global scale
in the sense defined above. It is determined by the expectation value
$a^2=\langle \frac{1}{N}Tr{\bf H}^2\rangle$. 
It is generally accepted to scale
 the entries $H_{ij}\sim 1/N^{1/2}$, i.e. 
in such a way that $a$ stays finite when $N\to \infty$,
the local spacing between eigenvalues in the neighbourhood of the
point $X$ being therefore $\Delta\propto 1/N$. 

From the point of view of 
universality the semicircular eigenvalue density
is not extremely robust.
Most easily one violates it by considering an
important class of the so-called "invariant ensembles"
characterized by a probability density
of the form ${\cal P}({\bf H})\propto
 \exp{-NTr V({\bf H})}$, with $V({\bf H})$ being an even polynomial.
The corresponding eigenvalue density turns out to be highly nonuniversal
and is determined by the particular form of the potential $V(H)$.
Only for $V({\bf H})={\bf H}^2$ it is given by the
semicircular law, Eq.(\ref{semi}). 
Actually, any ``deformation'' ${\bf H}_1={\bf H}_0+{\bf H}$ 
of the ensemble ${\bf H}$ with i.i.d. entries  by a given {\it fixed} 
matrix ${\bf H}_0$ results in the mean eigenvalue density  
belonging to a family of ``deformed semicircular laws'' 
discovered by Pastur\cite{Pas}.

Moreover, one can easily
 have a non-semicircular
eigenvalue density even for real symmetric matrices
${\bf S};\quad S_{ij}=S_{ji}$
with i.i.d. entries, if one keeps
the mean number of non-zero entries $p$ per column to be of the order
of unity when performing the limit $N\to\infty$.
This is a characteristic feature of the
so-called {\it sparse} random matrices\cite{MF,spa,FSC}
characterized by the following probability density of a given entry $S_{ij}$ :
\begin{equation}\label{distr}
{\cal P}(S_{ij})=(1-\frac{p}{N})\delta(S_{ij})+\frac{p}{N}h(S_{ij})
\end{equation}
where $h(s)=h(-s)$ is an arbitrary even distribution function satisfying the
conditions:
$h(0)<\infty;\quad \int h(s) s^{2} ds<\infty$.

Remarkably, a much more profound universality emerges for the
two-point correlation spectral function defined as:
\begin{equation}\label{cor}
\left\langle \rho(X_1)\rho(X_2)\right\rangle_c
=\left\langle
\rho(X_1)\right\rangle \delta(X_1-X_2)-{\cal Y}_2(X_1,X_2),
\end{equation}
where we defined the connected part of the correlation
function in a usual manner: $\langle AB\rangle_c=\langle AB\rangle-
\langle A\rangle\langle B\rangle$.
The nontrivial part of the spectral correlator is called
 the {\it cluster function} ${\cal Y}_2(X_1,X_2)$.
 It is one of the most informative statistical measures of the spectra,
\cite{Mehta}. It turns out, that already the global scale behaviour 
of ${\cal Y}_2(X_1,X_2)$
(i.e. one for the distance $S$ being comparable with the support of the
spectrum in the limit $N\to\infty$) is rather universal.
It is the same
for all the "invariant ensembles" \cite{univ1} and for those with i.i.d.
matrix elements\cite{kkp} and is
determined only by the positions of edge points $[A,B]$ of the
spectrum \cite{noteuni}.

Even more interesting is the fact that  universality of
 the correlation function Eq.(\ref{cor})
(as well as of all higher correlation functions) extends to the local scale,
i.e. for the distances $|X_1-X_2|$ comparable with
$\Delta$. This fact was rigorously 
proved for the unitary invariant ensembles\cite{Pas1}, extended to the
unitary  ``deformed'' ensembles\cite{BH} and heuristically verified
for other invariant ensembles\cite{HW}
as well as for the ensembles of sparse matrices\cite{MF,note}.
The particular form of the cluster function is different from that
typical for the global scale. It
is dictated in general by {\it global symmetries}
of the random matrices, e.g.if they are complex  Hermitian or real symmetric
\cite{Mehta}. All specific
(nonuniversal) properties are encoded in the value of local spacing $\Delta$. 
For the distances $S$ such that $S\gg\Delta$ local expressions
match with the global one, the latter taken at distances $S\ll [A-B]$.

It turns out, that it is the {\it local scale} universality
 that is mostly relevant for real physical systems\cite{Bohigas}.
Namely, the statistics of highly excited
bound states of {\it closed} quantum chaotic systems
of quite different microscopic nature turn out to be independent of
the microscopic details when sampled on the energy intervals large in
comparison with the mean level separation, but smaller
than the energy scale related by the Heisenberg uncertainty principle to the
relaxation time necessary for a classically chaotic system to reach
equilibrium in phase space \cite{AlSi}.
Moreover, the
spectral correlation functions turn out to
be exactly those which are provided by the theory of
large random matrices on the {\it local} scale \cite{per,MK},
with different symmetry classes
corresponding to presence or absence of the time-reversal symmetry.

Motivated by lack of general theoretical results
 for spectral properties of the pencils of random matrices, we 
start in Sec.II with considering an abstract 
pencil , with real symmetric ${\bf H}$ belonging to the 
Gaussian ensemble of random matrices with i.i.d. entries 
and $W$ being a real symmetric positive definite one 
with fixed given entries.
We derive an expression for the mean spectral density of such a pencil
and demonstrate that the correlation properties on the local scale
(in the sense defined above) are the same as given by 
the Wigner-Dyson expressions. Then, in Sec.III we extend our consideration
to the case of resonance statistics in the random infinite-range
LC-network, by deriving the mean density and the two-point spectral
correlation function for the resonances.

\section{Wigner-Dyson universality for spectra
of regular matrix pencils}

To study spectral properties of a regular matrix pencil
${\bf H}-\lambda {\bf W}$ we employ the Efetov supersymmetry approach
\cite{Efbook,VWZ}. A pedestrian introduction into the method 
can be found in \cite{my}. 

A convenient starting point is a representation of the spectral
density $\rho(\lambda)$ of real eigenvalues $\lambda_n$ 
for the equivalent symmetric eigenvalue problem 
for the matrix $\tilde{\bf H}=
{\bf W}^{-1/2}{\bf H}{\bf W}^{-1/2}$
in the following form:
\begin{equation}\label{dendef}
\rho(\lambda)=\frac{1}{N}\sum_{n=1}^N\delta(\lambda-\lambda_n)=
\mp\lim_{\epsilon\to +0}\frac{1}{\pi N} \im
\Tr\frac{1}{(\lambda\pm i\epsilon) {\bf 1}-\tilde{\bf H}}
\end{equation}
which after a trivial manipulation can be rewritten as
a derivative
\begin{equation}\label{deriv}
\rho(\lambda)=\mp\im \frac{1}{\pi N}\lim_{\epsilon\to +0}
\lim_{J\to 0}\frac{\partial}{\partial J} {\cal Z}_{\pm}(J,\lambda)
\end{equation}
of a generation function ${\cal Z}_{\pm}(J,\lambda)$ defined as
\begin{equation}\label{genfunc}
{\cal Z}_{\pm}(J,\lambda)=\frac{\det\left[(\lambda\pm i\epsilon+J){\bf
W}-{\bf H}\right]}{\det\left[(\lambda\pm i\epsilon){\bf
W}-{\bf H}\right]}
\end{equation}
In a completely analogous way one expresses the two-point
spectral correlation function ("cluster function", see Eq.(\ref{cor})):
\begin{equation}\label{tpdef}
-{\cal Y}_2(\lambda_1,\lambda_2)=\langle \rho(\lambda_1)\rho(\lambda_2)
\rangle-\langle \rho(\lambda_1)\rangle\langle\rho(\lambda_2)
\rangle=\frac{1}{2\pi^2}\re {\cal K}_{con}(\lambda_1,\lambda_2)
\end{equation}
as
\begin{equation}\label{tpgen}
{\cal K}_{con}(\lambda_1,\lambda_2)=\frac{1}{N^2}\lim_{\epsilon\to +0}
\lim_{J_{1,2}\to 0}\frac{\partial^2}{\partial J_1\partial J_2} 
\left[\langle {\cal Z}_{-}(J_1,\lambda_1){\cal Z}_{+}(J_2,\lambda_2)\rangle
-\langle {\cal Z}_{-}(J_1,\lambda_1)\rangle
\langle{\cal Z}_{+}(J_2,\lambda_2)\rangle\right]
\end{equation}
where the angular brackets stand for the ensemble averaging
and we assume that $\lambda_1\ne\lambda_2$.
The expressions above are actually valid in the limit $N\to \infty$, where
one can show that 
$$
\lim_{\epsilon\to +0}
\lim_{J_{1,2}\to 0}\frac{\partial^2}{\partial J_1\partial J_2} 
\left[\langle {\cal Z}_{\pm}(J_1,\lambda_1){\cal Z}_{\pm}(J_2,\lambda_2)\rangle
-\langle {\cal Z}_{\pm}(J_1,\lambda_1)\rangle
\langle{\cal Z}_{\pm}(J_2,\lambda_2)\rangle\right]=0
$$
and we used this fact in Eq.(\ref{tpdef}).

To facilitate the ensemble averaging we represent the ratio of the two  
determinants in Eq.(\ref{genfunc}) as the Gaussian integral
\begin{equation}\label{genf1}
{\cal Z}_{\pm}(J,\lambda)=(-1)^N\int \prod_{i=1}^{N}[d \Psi_{i}(\pm)] 
\exp\{\pm\frac{i}{2}\sum_{i,j}^N\Psi_i^{\dagger}(\pm)
\left[W_{ij}(\lambda\pm i\epsilon+J\hat{k})-H_{ij}\right]\Psi_j(\pm)\}
\end{equation}
over 4-component supervectors $\Psi_{i}(\pm)$,
\begin{equation}\label{suvec1}
\Psi_{i}(\pm)=
\left(
\begin{array}{c} 
\vec{R}_{i}(\pm)\\
\vec{\eta}_{i}(\pm)
\end{array}
\right),
\vec{R}_{i}(\pm)=
\left(
\begin{array}{c}
r_{i}(\pm)\\
r_{i}^{*}(\pm)
\end{array}
\right),
\vec{\eta}_{i}(\pm)=
\left(
\begin{array}{c}
\chi_{i}(\pm)\\ 
\chi_{i}^{*}(\pm)
\end{array}
\right),
d\Psi_i=\frac{dr_idr_i^*}{2\pi}d\chi_i^*d\chi_i
\end{equation}
with the components $r_{i}(+),r_{i}(-);\quad i=1,2,...,N$ 
being complex commuting
variables and $\chi_{i}(+),\chi_{i}(-)$ forming the 
corresponding Grassmannian
parts of the supervectors $\Psi_{i}(\pm)$.
A $4\times 4$ diagonal supermatrix $\hat{k}=\diag(0,0,1,1)$
takes care of absence of the source $J$ in the denomenator of
the generating function Eq(\ref{genfunc}).

Since we are dealing finally with the averaged product of two
such generating functions it is convenient
 to introduce the $8-$component supervectors
\begin{equation}\label{suvec2}
\Phi_{i}=
\left(
\begin{array}{c}
\Psi_{i}(+)\\ 
\Psi_{i}(-)
\end{array}
\right)
\end{equation}
and finally $\Phi=\left(\Phi_1,...,\Phi_N\right)^T$ as well as the
supermatrices $\hat{\Lambda}=\diag(1,1,1,1,-1,-1,-1,-1)$ and 
$\hat{J}_k=\diag(J_{1}\hat{k},J_{2}\hat{k})$.

It is also useful to remember that we expect to have non-trivial
spectral correlations on the scale comparable with the mean 
spacing between neigbouring eigenvalues, i.e. when 
$\lambda_1-\lambda_2\propto 1/N$. Correspondingly, we introduce:
$\lambda_1=\lambda-s/2N\quad,\quad \lambda_2=\lambda+s/2N$
and consider $s$ to be of the order of unity henceforth.
All this allows us to write the product of two generating functions
as
\begin{eqnarray}\label{genprod}
{\cal Z}^{\lambda,s}(J_1,J_2)=
{\cal Z}_{-}(J_1,\lambda_1){\cal Z}_{+}(J_2,\lambda_2)=\\ \nonumber
\int d\Phi\exp\left\{-\frac{i}{2}\Phi^{\dagger}\hat{\Lambda}\otimes
{\bf H}\Phi + \frac{i}{2}\Phi^{\dagger}\left[\lambda \hat{\Lambda}
-(\epsilon-is/2N)\hat{1}+\hat{J}_k\right]\otimes {\bf W}\Phi\right\}
\end{eqnarray}

This expression can be easily averaged over the Gaussian distribution
of ${\bf H}$ by the chain of identities:
\begin{equation}\label{HST}
\left\langle e^{-\frac{i}{2}\Phi^{\dagger}\hat{\Lambda}\otimes
{\bf H}\Phi}\right\rangle=e^{-\frac{a^2}{4N}\Str \hat{A}^2}=
\int d\hat{Q}\exp{\left[-\frac{N}{4}\Str\hat{Q}^2-i\frac{a}{2}
\Phi^{\dagger}\hat{\Lambda}\hat{Q}\Phi\right]}
\end{equation}
where $a$ determines the variance of matrix elements of ${\bf H}$,
see Eq.(\ref{semi})
and $\hat{A}=\hat{\Lambda}^{1/2}
\left(\sum_i \Phi_i\otimes\Phi^{\dagger}\right)\hat{\Lambda}^{1/2}$.
The last relation trading the term  in the exponent
quartic with respect to $\Phi$ for an auxilliary
integration over the set of supermatrices $\hat{Q}$ is known as
the Hubbard-Stratonovich transformation and plays a cornerstone
role in the whole method. After substituting Eq.(\ref{HST})
back into averaged Eq.(\ref{genprod}) and changing the order of integrations
one performs the (Gaussian) integral over $\Phi$ explicitly.
It turns out that in order to justify all these operations 
formally one has
to restrict the supermatrices $\hat{Q}$ to a manifold paramatrized
as $\hat{Q}=\hat{T}^{-1}\hat{P}\hat{T}$, with $\hat{P}$ being
a block-diagonal Hermitian supermatrix and $\hat{T}$ belonging to a 
certain graded coset space $UOSP(2,2/4)/UOSP(2/2)\otimes UOSP(2/2)$.
A detailed discussion of this fact and an explicit parametrisation of
the matrices $\hat{T}$ could be found in \cite{Efbook,VWZ,my,EAI}.

The resulting expression for the averaged product
of the generation functions turns out to be dependent only on
eigenvalues $w_i\,,\,i=1,...,N$ of the matrix ${\bf W}$
(this fact can be traced back to the "rotational invariance" of the
Gaussian Orthogonal Ensemble (GOE) formed by ${\bf H}$) 
and has the following form:
\begin{equation}\label{genave}
\langle{\cal Z}^{\lambda,s}(J_1,J_2) \rangle=
\int d\hat{Q}\exp{\left[-\frac{N}{4}\Str\hat{Q}^2-\frac{1}{2}
\sum_{i=1}^N\Str\ln\left\{\left(\lambda\hat{1}-v_i\hat{Q}\right)+
\left(i\epsilon+s/2N+\hat{J}_k\right)\hat{\Lambda}\right\}\right]}
\end{equation}
where $v_i=a/w_i$.

Since we are interested in the limit $(\epsilon,J,1/N)\to 0$, we 
can expand the logarithm in the exponent correspondingly:
\begin{equation}\label{e0}
\begin{array}{c} 
\Str\ln\left\{\left(\lambda\hat{1}-v_i\hat{Q}\right)+
\left(i\epsilon+s/2N+\hat{J}_k\right)\hat{\Lambda}\right\}=\\
\Str\ln\left(\lambda\hat{1}-v_i\hat{Q}\right)
+\Str\left(\lambda\hat{1}-v_i\hat{Q}\right)^{-1}\hat{\Lambda}
\left(i\epsilon+s/2N+\hat{J}_k\right)-\frac{1}{2}
\Str\left[\left(\lambda\hat{1}-v_i\hat{Q}\right)^{-1}\hat{\Lambda}
\hat{J}_k\right]^2+...
\end{array}
\end{equation}
so that differentiating the exponent over the sources $J_{1,2}$ yields
the preexponential factors:
\begin{equation}\label{e1}
\lim_{J_{1,2}\to 0}\frac{\partial}{\partial J_1}\exp\{...\}\propto
\sum_{i=1}^N\Str
\left[\left(\lambda\hat{1}-v_i\hat{Q}\right)^{-1}\hat{k}
\frac{\hat{\Lambda}+\hat{1}}{2}\right]
\end{equation}
and
\begin{equation}\label{e2}
\begin{array}{c}
\lim_{J_{1,2}\to 0}\frac{\partial^2}{\partial J_1\partial J_2}
\exp\{...\}\propto 
2\sum_{i=1}^N\Str\left[\left(\lambda\hat{1}-v_i\hat{Q}\right)^{-1}\hat{k}
\frac{\hat{\Lambda}+\hat{1}}{2}\left(\lambda\hat{1}-v_i\hat{Q}\right)^{-1}
\hat{k}\frac{\hat{\Lambda}-\hat{1}}{2}\right]\\
+\sum_{i=1}^N\Str\left[\left(\lambda\hat{1}-v_i\hat{Q}\right)^{-1}\hat{k}
\frac{\hat{\Lambda}+\hat{1}}{2}\right]
\sum_{i=1}^N\Str\left[
\left(\lambda\hat{1}-v_i\hat{Q}\right)^{-1}
\hat{k}\frac{\hat{\Lambda}-\hat{1}}{2}\right]
\end{array}
\end{equation}
In the limit $N\gg 1$ each sum over $i$ is of the order of $N$, so that
a contribution of second term in the expression above is by factor $N$ larger
than the contribution of the first one.
In other words, we could restrict ourselves 
by terms linear in sources $J_{1,2}$ in the expansion Eq.(\ref{e0})
above. We will make use of this fact later on when considering 
 a more complicated situation in the next section.

Taking all this into account, we arrive at the following 
integral representation:
\begin{eqnarray}\label{tprep}
\nonumber
{\cal K}(\lambda_1,\lambda_2)&=&\frac{1}{N^2}\lim_{\epsilon\to +0}
\lim_{J_{1,2}\to 0}\frac{\partial^2}{\partial J_1\partial J_2} 
\langle {\cal Z}_{-}(J_1,\lambda_1)
{\cal Z}_{+}(J_2,\lambda_2)\rangle\\
&=&\frac{1}{4N^2}\lim{\epsilon\to 0}\int dQ 
\exp{\left[-N{\cal L}(\hat{Q})-\frac{i}{2}
\left(\epsilon-is/2N\right)\sum_i
\Str\left(\lambda\hat{1}-v_i\hat{Q}\right)^{-1}\hat{\Lambda}\right]}
\\ \nonumber && \times
\sum_{i=1}^N\Str\left[\left(\lambda\hat{1}-v_i\hat{Q}\right)^{-1}\hat{k}
\frac{\hat{\Lambda}+\hat{1}}{2}\right]
\sum_{i=1}^N\Str\left[
\left(\lambda\hat{1}-v_i\hat{Q}\right)^{-1}
\hat{k}\frac{\hat{\Lambda}-\hat{1}}{2}\right]
\end{eqnarray}
where 
\begin{equation}\label{act}
{\cal L}(\hat{Q})=\frac{1}{2}\Str \hat{Q}^2+\frac{1}{N}
\sum_i\Str\ln\left(\lambda\hat{1}-v_i\hat{Q}\right)
\end{equation}

In the same way one obtains the expression for the mean spectral
density:
\begin{equation}\label{meand}
\rho(\lambda)=\frac{1}{N\pi}
\lim_{\epsilon \to 0}\im\int dQ 
\sum_{i=1}^N\Str\left[\left(\lambda\hat{1}-v_i\hat{Q}\right)^{-1}\hat{k}
\frac{\hat{\Lambda}+\hat{1}}{2}\right]
\exp{\left[-N{\cal L}(\hat{Q})-\frac{i}{2}
\epsilon \sum_i
\Str\left(\lambda\hat{1}-v_i\hat{Q}\right)^{-1}\hat{\Lambda}\right]}
\end{equation}

In the limit $N\to \infty$ the integrals over $\hat{Q}$ in the
expressions above are dominated by saddle points of the "action"
${\cal L}(\hat{Q})$ Eq.(\ref{act}) which satisfy the equation:
\begin{equation}\label{sp}
\hat{Q}=\frac{1}{N}\sum_{i=1}^N\frac{v_i}{\lambda\hat{1}-v_i\hat{Q}}
\end{equation}
relevant solutions of which belonging to the 
integration domains $\hat{Q}=\hat{T}^{-1}\hat{P}\hat{T}$ are parametrized as
follows:
\begin{equation}\label{spm}
\hat{Q}_{s.p.}=t\hat{1}+iq\hat{T}^{-1}\hat{\Lambda}\hat{T}
\end{equation}
where the real parameters $t,q$ satisfy for $q\ne 0$ the system of
two equations:
\begin{equation}\label{sys}
t=\frac{1}{N}\sum_i\frac{v_i(\lambda-v_it)}{(\lambda-v_it)^2+q^2v_i^2}
\quad \mbox{and} \quad 
1=\frac{1}{N}\sum_i\frac{v_i^2}{(\lambda-v_it)^2+q^2v_i^2}
\end{equation}
Let us note, that the saddle-point solutions Eq.(\ref{spm}) form
for $q\ne 0$ a continuos manifold 
parametrized by the supermatrices $\hat{T}$.

Using these expressions it is easy to invert the matrix
$\left(\lambda\hat{1}-v_i\hat{Q}_{s.p}\right)$ and
check that
\begin{eqnarray}\label{pree}
\sum_{i=1}^N\Str\left[\left(\lambda\hat{1}-v_i\hat{Q}_{s.p}\right)^{-1}\hat{k}
\frac{\hat{\Lambda}\pm\hat{1}}{2}\right] &=& 
\pm 2 \frac{\lambda-v_it}{(\lambda-v_it)^2+q^2v_i^2}+
iq\frac{v_i}{(\lambda-v_it)^2+q^2v_i^2}
\Str\left[\hat{T}^{-1}\hat{\Lambda}\hat{T}\hat{k}
\frac{\hat{1}\pm\hat{\Lambda}}{2}\right]\\
\sum_i\Str\left(\lambda\hat{1}-v_i\hat{Q}_{s.p}\right)^{-1}\hat{\Lambda}&=&
i\Str\left[\hat{T}^{-1}\hat{\Lambda}\hat{T}\hat{\Lambda}\right]
\sum_i\frac{v_i q}{(\lambda-v_it)^2+q^2v_i^2}
\end{eqnarray}

So the problem amounts to substituting these expressions
into the integrand of Eqs.(\ref{tprep},\ref{meand})
and to performing the remaining integration over
the coset space parametrized by the matrices $\hat{T}$.
The way of doing it described in much detail in \cite{Efbook,VWZ,my,EAI}
and we refer the interested reader to those papers.

Here we mention only a few most important moments.
First of all the integral in Eq.(\ref{meand})
is given by the so-called PSEW
(Parisi-Sourlas-Efetov-Wegner) theorem
 due to a specific symmetry of the integrand (
in the limit $\epsilon\to 0$ the integrand contains only a part of the integration
variables due to the projector $(\hat{\Lambda}+1)/2$)
The resulting mean eigenvalue density is merely given by:
\begin{equation}\label{denav}
\rho(\lambda)=\frac{q}{\pi
N}\sum_i\frac{v_i}{(\lambda-v_it)^2+q^2v_i^2} 
\end{equation}
The system of equations Eqs(\ref{sys},\ref{denav})
turns out to be in fact equivalent to a particular case of general
results by Pastur and Girko\cite{Pas}.

Substituting the expression Eq.(\ref{denav}) for the mean spectral
density to Eqs.(\ref{pree}) and the resulting formulae further
to the integral Eq.(\ref{tprep},\ref{act}) 
for the averaged generating function
 one can express the 
"connected" part of the two-point correlation function 
${\cal K}_{con}(\lambda_1,\lambda_2)$ as: 
\begin{eqnarray}\label{con}
{\cal K}_{con}(\lambda_1,\lambda_2)&=&\\ \nonumber
&\frac{\pi^2\rho^2(\lambda)}{4}&
\int d\mu{(T)}\Str\left[\hat{T}^{-1}\hat{\Lambda}\hat{T}\hat{k}
\frac{\hat{1}+\hat{\Lambda}}{2}\right]
\Str\left[\hat{T}^{-1}\hat{\Lambda}\hat{T}\hat{k}
\frac{\hat{1}-\hat{\Lambda}}{2}\right]
\exp\left\{-\frac{i\pi \rho(\lambda)}{4}s
\Str\left[\hat{T}^{-1}\hat{\Lambda}\hat{T}\hat{\Lambda}\right]
\right\}
\end{eqnarray}
with $d\mu(T)$ being an appropriate invariant measure on the coset space.
When deriving that expression from
Eqs.(\ref{tprep},\ref{act},\ref{pree})  we noticed that
the "disconnected" part of the correlation function
is again given by PSEW theorem and cancels exactly
the contribution from those few terms in the integrand which
are not proportional to the mean density $\rho(\lambda)$.

The expression Eq.(\ref{con}) is our main result and 
is quite remarkable: its form 
exactly coincides with the corresponding expression for the
two-point correlation function  of random matrices from GOE, 
see e.g. \cite{EAI}. This fact was first demonstrated by
Efetov who managed to perform a non-trivial
integration over the coset-space explicitly\cite{Efbook} and found
that it reproduced the famous Dyson expression\cite{notcor} for the 
two-point function\cite{Mehta}:
\begin{equation}\label{Dyson}
{\cal Y}_2(\lambda_1,\lambda_2)=\left(\frac{\sin s_e}{s_e}\right)^2+
\frac{d}{d s_e}\left(\frac{\sin s_e}{s_e}\right)
\int_1^{\infty}dt \frac{\sin{(s_et)}}{t}
\end{equation}
where $s_e=\pi\rho(\lambda)s$ is a spectral distance $s/N$ measured in
units of the local mean spacing $\Delta=1/[N\rho(\lambda)]$.

In the end of the present section we just mention that our
method of deriving Eq.(\ref{con}) can be easily generalized
to matrix pencils whose random part ${\bf H}$ is an arbitrary
matrix (real symmetric or Hermitian) with i.i.d. entries
 following the lines of papers\cite{MF,FSC}. In the same way
one can show that all higher correlation functions also coincide
with GOE expressions.

\section{Spectral properties of a "full connectivity" LC network}

Now we are well-prepared to calculate the mean density of resonances and
 the corresponding two-point spectral correlation function of the "full-connectivity"
LC network as defined by eqs.(\ref{defin},\ref{reson}). Actually, it is convenient to rescale both matrices
in Eq.\ref{defin} as: ${\bf H}\to \frac{1}{N^{1/2}}{\bf H},\quad
{\bf W}\to \frac{1}{N^{1/2}}{\bf W}$. This transformation obviously does
not change generalized eigenvalues of the pencil 
${\bf H}-\tilde{\lambda}{\bf W}$
but facilitates bookkeeping of the leading terms in various expansions.

In our consideration we follow the pattern of the previous section
and introduce the generation function identical to Eq.(\ref{genprod}):
\begin{eqnarray}\label{gef}
{\cal Z}^{r,s}(J_1,J_2)=\int \prod_i d\Phi_i 
\exp\left\{-\frac{i}{2N^{1/2}}\sum_{i=1}^N\Phi_i^{\dagger}
\left[(h_{Ai}+h_{Bi})\hat{\Lambda}-2i\tilde{\epsilon}\hat{1}
-2\left(\frac{r}{N^{1/2}}\hat{\Lambda}+\hat{J}\right)\right]\Phi_i\right\}\\
\nonumber \times 
\exp\left\{\frac{i}{2N^{1/2}}\sum_{i<j}^N
\left(\Phi_i^{\dagger}-\Phi_j^{\dagger}\right)
\left[\frac{r}{N^{1/2}}\hat{\Lambda}+\hat{J}_k+i\tilde{\epsilon}\hat{1}
+h_{ij}\hat{\Lambda}\right]\left(\Phi_i-\Phi_j\right)\right\}
\end{eqnarray}
where we used that 
\[
\frac{1}{2}\sum_{i<j}h_{ij}(\Phi_i^{\dagger}-\Phi_j^{\dagger})\hat{B}
(\Phi_i-\Phi_j)=-\sum_{i<j}h_{ij}\Phi_i^{\dagger}\hat{B}\Phi_j+
\frac{1}{2}\sum_i\Phi_i^{\dagger}\hat{B}\Phi_i\left(\sum_{k\ne
i}h_{ik}\right) 
\]
for any supermatrix $\hat{B}$ as long as
$\Phi_i^{\dagger}\hat{B}\Phi_j=\Phi_j^{\dagger}\hat{B}\Phi_i$. 
We also  envisaged that for the full-connectivity 
network the resonance density is
nonvanishing as long as $\tilde{\lambda}\sim 1/N^{1/2}$ and 
introduced the scaled variable $r=\tilde{\lambda} N^{1/2}$ (see
the Introduction). Then the typical spacing between neighbouring
resonances should be of the order of $\Delta\sim N^{-3/2}$. Since we can 
expect a nontrivial spectral correlation on the distances 
$S=\tilde{\lambda}_2-\tilde{\lambda}_1$ 
comparable with $\Delta$ we introduced the quantity 
$\tilde{\epsilon}=\epsilon-is/(2N^{3/2})$ and consider $s$ to be of
the order of unity. All other notations coincide with those in 
Eq.(\ref{genprod}).

Now we have to perform the averaging over 
$h_{\mu\nu},\,\, (\mu,\nu)=A,B,1,...,N$, each taking values $\pm 1$
with the equal probabilities.
This is done in the limit $N\to \infty$ as follows:
\begin{eqnarray}\label{avpr}\nonumber
&&\prod_{i<j}\left\langle \exp\left\{\frac{i}{2N^{1/2}}h_{ij}
\left(\Phi_i^{\dagger}-\Phi_j^{\dagger}\right)
\hat{B}\left(\Phi_i-\Phi_j\right)\right\}\right\rangle=\\ \nonumber
&&\prod_{i<j}\frac{1}{2}\left[
\exp\left\{\frac{i}{2N^{1/2}}
\left(\Phi_i^{\dagger}-\Phi_j^{\dagger}\right)
\hat{B}\left(\Phi_i-\Phi_j\right)\right\}+
\exp\left\{-\frac{i}{2N^{1/2}}
\left(\Phi_i^{\dagger}-\Phi_j^{\dagger}\right)
\hat{B}\left(\Phi_i-\Phi_j\right)\right\}
\right]=\\
&&\prod_{i<j}\left\{1-\frac{1}{8N}\left[
\left(\Phi_i^{\dagger}-\Phi_j^{\dagger}\right)
\hat{B}\left(\Phi_i-\Phi_j\right)\right]^2+
...\right\}\approx \exp\{-\frac{1}{16N}
\sum_{i,j}\left[
\left(\Phi_i^{\dagger}-\Phi_j^{\dagger}\right)
\hat{B}\left(\Phi_i-\Phi_j\right)\right]^2\}
\end{eqnarray}
and in similar way we can average also over $h_{A,i},h_{B,i}$.
Introducing now the notations:
\begin{eqnarray}\label{kern}
{\cal K}(\Phi_a,\Phi_b)=-4ir\left(\Phi_a^{\dagger}-\Phi_b^{\dagger}\right)
\hat{\Lambda}\left(\Phi_a-\Phi_b\right)
+\left[ \left(\Phi_a^{\dagger}-\Phi_b^{\dagger}\right)
\hat{\Lambda}\left(\Phi_a-\Phi_b\right)\right]^2\\
f(\Phi)=\frac{i}{N^{1/2}}\Phi^{\dagger}\left(i\tilde{\epsilon}\hat{1}
+\frac{r}{N^{1/2}}\hat{\Lambda}+\hat{J}_k\right)\Phi-\frac{1}{4N}
\left(\Phi^{\dagger}\hat{\Lambda}\Phi\right)^2
\label{f}
\end{eqnarray}
we can rewrite the averaged generating function as:
\begin{eqnarray}\label{genaver}
{\cal Z}(J_1,J_2)_{av}&=&\int \prod_i d\Phi_i
\exp\left\{\sum_i f(\Phi_i)-
\frac{1}{16N}\sum_{ij}{\cal K}(\Phi_i,\Phi_j)\right\}\\ \nonumber
&&\times\exp\left\{\frac{i}{2N^{1/2}}\sum_{i<j}
\left(\Phi_i^{\dagger}-\Phi_j^{\dagger}\right)
\left[i\tilde{\epsilon}\hat{1}+\hat{J}_k\right]\left(\Phi_i-\Phi_j\right)
\right\}
\end{eqnarray}

For the case of Gaussian real symmetric matrices with independent entries
 ${\bf H}$ considered in 
the previous section, the key point allowing us to make progress was the
exploitation of the Hubbard-Stratonovich decoupling; see Eq.(\ref{HST}).  
In a similar way, we proceed here with using a 
{\it functional generalization} of the Hubbard-Stratonovich identity
suggested by us earlier\cite{MF} in the context of studies of the sparse
random matrices:
\begin{equation}\label{FHTS}
\exp\left\{-\frac{1}{16N}\sum_{i,j}{\cal K}(\Phi_i,\Phi_j)\right\}=
\int{\cal D}(g)\exp\left\{-\frac{N}{16}\int d\Phi_ad\Phi_b
g(\Phi_a){\cal C}(\Phi_a,\Phi_b)g(\Phi_b)+
\frac{i}{8}\sum_{i=1}^N g(\Phi_i)\right\}
\end{equation}
where the kernel ${\cal C}(\Phi_a,\Phi_b)$ is , in a sense, 
the inverse of a (symmetric) kernel ${\cal K}(\Phi_a,\Phi_b)$:
\begin{equation}\label{inv}
\int d\Psi {\cal K}(\Phi_a,\Psi){\cal
C}(\Psi,\Phi_b)=\delta(\Phi_a,\Phi_b) 
\end{equation}
and $\delta(\Phi_a,\Phi_b)$ plays the role of a $\delta-$ functional
kernel in a space spanned by the functions $g(\Phi)$. Some hints towards
understanding of the identities Eq.(\ref{FHTS},\ref{inv}) 
 are given in the Appendix.

With the help of these relations one easily brings the averaged generation
function eq.(\ref{genaver}) to the form:
\begin{eqnarray}\label{zdec}
\langle{\cal Z}^{r,s}(J_1,J_2)\rangle&=&\int {\cal D}(g)
\exp\{-N{\cal L}(g)+\delta{\cal L}_1(g)\}\\
{\cal L}(g)&=&\frac{1}{16}\int d\Phi_ad\Phi_b
g(\Phi_a){\cal C}(\Phi_a,\Phi_b)g(\Phi_b)-
\ln{\int d\Phi e^{\frac{i}{8}g(\Phi)+f(\Phi)}}\\
\delta{\cal L}_1(g)&=&\ln\left[
\frac{\int \prod_id\Phi_i\exp\left\{\sum_i\left(\frac{i}{8}
g(\Phi_i)+f(\Phi_i)\right)+\frac{i}{2N^{1/2}}\sum_{i<j}
\left(\Phi_i^{\dagger}-\Phi_j^{\dagger}\right)
\left[i\tilde{\epsilon}\hat{1}+\hat{J}_k\right]\left(\Phi_i-\Phi_j\right)
\right\}}
{\int \prod_id\Phi_i\exp{\sum_i\left(\frac{i}{8}
g(\Phi_i)+f(\Phi_i)\right)}}\right] \label{zdec1}
\end{eqnarray}
As is clear from these expressions, the term ${\cal L}_1(g)$
is only a small correction to the main term in the exponent
of the functional integral when $N\to \infty, \quad J,\epsilon\to 0$.
Therefore, the functional integration over $g$ can be performed
by the saddle-point method (cf.Eqs.(\ref{tprep},\ref{meand})). The saddle-point 
configuration $g_s(\Phi)$
can be found by requiring the vanishing variation of the "action"
${\cal L}(g)$ and satisfies the following equation:
\begin{equation}\label{fsp}
g(\Phi_a)=i\frac{\int d\Phi_b 
{\cal K}(\Phi_a,\Phi_b)e^{\frac{i}{8}g(\Phi_b)+f(\Phi_b)}}
{\int d\Phi_b e^{\frac{i}{8}g(\Phi_b)+f(\Phi_b)}}
\end{equation}
When deriving Eq.(\ref{fsp}) we have used Eq.(\ref{inv}).
It is completely clear that in the limit $N\to \infty$ one can safely
disregard the term $f(\Phi_b)$, Eq.(\ref{f}) 
as being of the next order in $1/N$
in comparison with $g(\Phi_b)$.
From now on we just put $f=0$ everywhere.

Given the form of the kernel Eq.(\ref{kern}), we managed to guess 
 the following explicit solution to the saddle-point equation:
\begin{equation}\label{ansatz}
g_s(\Phi_a)=4(r-G_1)(\Phi_a^{\dagger}\hat{\Lambda}\Phi_a)
+4iG_2(\Phi_a^{\dagger}\Phi_a)+i(\Phi_a^{\dagger}\hat{\Lambda}\Phi_a)^2
\end{equation}
Substituting such an Ansatz to Eq.(\ref{fsp}) one can verify by
a direct calculation that it indeed satisfies the saddle-point equation
provided the real coefficients $G_1,G_2$ are solutions of the system
of two conjugate equations:
\begin{eqnarray}\label{system}
G_2+iG_1&=&\int_0^{\infty}du
\exp\left\{\frac{i}{2}u(r-G_1+iG_2)-\frac{u^2}{8}\right\} \\ \nonumber
G_2-iG_1&=&\int_0^{\infty}du
\exp\left\{-\frac{i}{2}u(r-G_1-iG_2)-\frac{u^2}{8}\right\}
\end{eqnarray}
The following few identities might be helpful when 
performing a verification of such a fact.
Suppose we have a $4-$ component supervector $\Psi$ like that defined
in Eq.(\ref{suvec1}) and consider a function 
$F(\Psi)=\tilde{F}(\Psi^{\dagger}\Psi)$ vanishing on the 
boundary of integration. Then it is easy to verify that:
\begin{equation}
\int d\Psi F(\Psi)=\tilde{F}(0)\quad,\quad 
\int d\Psi_b (\Psi_a^{\dagger}\Psi_b) F(\Psi_b)=0
\end{equation}
the first identity being just a particular case of the PSEW theorem
mentioned in the previous section, see e.g.\cite{my}.

For further analysis it is very important that 
a solution for the system Eq.(\ref{system})
 exists for arbitrary $-\infty<r<\infty$
such that $G_2(r)>0$. We will see later on that the mean density of 
resonances is merely given by $\rho(r)=\frac{1}{\pi}G_2(r)$.
For $|r|\gg 1$ one can easily infer from Eqs.(\ref{system})
that $G_2(r)$ developes a gaussian tail: 
\[
G_2(r\gg 1)\approx (\pi/2)^{1/2}\exp\{-\frac{r^2}{2}\}
\] 
It is necessary to mention that the equations Eq.(\ref{system})
virtually coincide with those emerging in a study of density 
of eigenvalues of a transition matrix on a randomly diluted 
graph in a limit of 
large connectivity performed by Bray and Rodgers\cite{BRD}, see
their Eqs.(20),(21) and fig.1 in their paper.
Indeed two problems have many common features, and 
the coincidence is hardly accidental. An indication of a relation
between the problems comes from the structure of the matrix ${\bf \hat{H}}$, 
see eq.(\ref{defin}), which is actually very similar to the structure of the matrix considered by Bray and Rodgers. Moreover, 
the resonances we study are eigenvalues of the matrix ${\bf \tilde{H}}={\bf W}^{-1/2}
{\bf \hat{H}} {\bf W}^{-1/2}$, and it is easy to write down explicit expressions for the entries of ${\bf \tilde{H}}$ because of the simple structure
of the matrix ${\bf W}$, see eq.(\ref{defin}). One finds that ${\bf{\tilde{H}}}=
\frac{1}{N}{\bf H}+O(1/N^{3/2})$ which provides a direct link to the paper
\cite{BRD} in the limit $N\to \infty$.

The most important consequence of the existence of such a solution
$g_s(\Phi_a)$, see Eq.(\ref{ansatz}), with $G_2\ne 0$  is actually the
simultaneous existence of a whole {\it continous manifold} of the saddle point
solutions parametrized as:
\begin{equation}
g_T(\Phi_a)=g_s(\hat{T}\Phi_a),\quad \mbox{with} 
\quad \hat{T}^{\dagger}\hat{\Lambda}\hat{T}=\hat{\Lambda}
\end{equation}
Indeed, from the invariance property of the kernel Eq.(\ref{kern}):
${\cal K}(\Phi_a,\Phi_b)={\cal K}(\hat{T}\Phi_a,\hat{T}\Phi_b)$
and from the (pseudo)unitarity of the matrices $T$:
$|\mbox{Sdet}\hat{T}|=1$ it follows that $g_s(\hat{T}\Phi_a)$ must be a 
solution to Eq.(\ref{fsp})
together with any given solution $g_s(\Phi_a)$. However, if it hadn't been
for  the condition $G_2\ne 0$ all these solutions would trivially 
coincide: $g_T(\Phi)\equiv g(\Phi)$ for
any $\hat{T}$ defined as above, i.e. the symmetry of the solution
would coincide with the symmetry of the equation itself.

In the actual case $G_2\ne 0$  presence of the combination
$\Phi_a^{\dagger}\Phi_a$  which is {\it not invariant} with respect  
to a transformation $\Phi_a\to \hat{T}\Phi_a$ makes
the symmetry of the solution Eq.(\ref{ansatz})
to be lower than the symmetry of the
equation Eq.(\ref{fsp}). This is an example of a very well-known effect 
of spontaneous breakdown of symmetry. 
The phenomenon makes the situation to be less 
trivial and generates the whole 
manifold of the saddle-point solutions. Different nontrivial solutions
are actually parametrized by the supermatrices $\hat{T}$ which are
elements of the same\cite{notegrad} graded coset space $UOSP(2,2/4)/UOSP(2/2)\otimes 
UOSP(2/2)$ which already appeared in the previous section,
see a discussion after Eq.(\ref{HST}).

It is simple to satisfy oneself that ${\cal L}(g_s)=0$, and 
the same obviosly holds for the whole manifold $g_T$. Therefore,
the only term that renders the expression for the generation function
to be non-trivial is $\delta{\cal L}_1(g_T)$. In the limit $(\epsilon 
,J,1/N)\to 0$ we can expand Eq.(\ref{zdec1}) as:
\begin{eqnarray}\label{expan}  
\delta{\cal L}_1(g_T)=\ln\left\{1+\frac{i}{2N^{1/2}}
\sum_{i<j}\int d\Phi_i d\Phi_j 
\left(\Phi_i-\Phi_j\right)^{\dagger}
\left(i\tilde{\epsilon}\hat{1}+\hat{J}\right)
\left(\Phi_i-\Phi_j\right)
e^{\frac{i}{8}\left(g_T(\Phi_i)+g_T(\Phi_j)\right)}\right\}\\
\nonumber
\approx \frac{i}{2}N^{3/2}\int d\Phi\,\, \Phi^{\dagger}
\left(i\tilde{\epsilon}\hat{1}+\hat{J}\right)\Phi\,\,
e^{\frac{i}{8}g_T(\Phi)}=\frac{i}{2}N^{3/2}\Str\left[\hat{W}
\left(i\tilde{\epsilon}\hat{1}+\hat{J}\right)\right]
\end{eqnarray}
where we introduced a supermatrix:
\begin{equation}\label{Wdef}
W_{\alpha\beta}=\int d\Phi \,\Phi_{\alpha}\Phi^{\dagger}_{\beta}
e^{\frac{i}{8}g_T(\Phi)}
\end{equation}
and used that fact, that it is enough for our purposes to keep
only terms linear with respect to the source matrix $\hat{J}$ (see the discussion after Eq.(\ref{e2})).

To determine the matrix elements of $\hat{W}$ it is convenient to use
the equation:
\begin{equation}\label{Wsearch1}
G_2\Phi_a^{\dagger}\Phi_a+iG_1\Phi_a^{\dagger}\hat{\Lambda}\Phi_a
=\int d\Phi_b 
\left(\Phi_{a}^{\dagger}\hat{\Lambda}\Phi_b\right)
\left(\Phi^{\dagger}_{b}\hat{\Lambda}\Phi_a\right)
e^{\frac{i}{8}g_s(\Phi_b)}
\end{equation}
which follows from the saddle-point equation Eq.(\ref{fsp})
and Eq.(\ref{ansatz}). Now we make a transformation:
$\Phi_{a,b}\to \hat{T}\Phi_{a,b}$ and easily find that:
\begin{eqnarray}\label{Wsearch2}
&&G_2\Phi_a^{\dagger}\hat{T}^{\dagger}\hat{T}\Phi_a+
iG_1\Phi_a^{\dagger}\hat{\Lambda}\Phi_a
=\int d\Phi_b 
\left(\Phi_{a}^{\dagger}\hat{\Lambda}\Phi_b\right)
\left(\Phi^{\dagger}_{b}\hat{\Lambda}\Phi_a\right)
e^{\frac{i}{8}g_T(\Phi_b)}\\
\nonumber
&=& \sum_{\alpha,\beta=1}^8\Phi_{a,\alpha}\Phi_{a,\beta}^{\dagger}
\Lambda_{\alpha\alpha}\Lambda_{\beta\beta}K_{\alpha\alpha}
W_{\beta\alpha}
\end{eqnarray}
where $K_{\alpha\alpha}=-1$ for $\alpha$ being a "fermionic" index
and unity otherwise.  Comparing both sides of that relation
(i.e. the coefficients of
 the quadratic forms with respect to the components of the supervector
$\Phi_{a}$) yields a relation between  $W$, the functions $G_{1,2}$
and the matrices $\hat{T}$:
\begin{equation}\label{W}
\hat{W}=G_2
\hat{T}^{-1}\hat{\Lambda}\hat{T}\hat{\Lambda}+iG_1\hat{\Lambda} 
\end{equation}
where we used $\hat{T}^{\dagger}\hat{T}=
\Lambda\hat{T}^{-1}\hat{\Lambda}\hat{T}$.

Using this fact we find that the averaged generating function
is expressed as an integral over the saddle-point manifold
parametrized by the matrices $\hat{T}$:
\begin{eqnarray}\label{genaverT}
\langle{\cal Z}^{r,s}(J_1,J_2)\rangle=\int d\mu(T)
\exp\left\{i\frac{s}{4}G_2(r)
\Str\left(\hat{T}^{-1}\hat{\Lambda}\hat{T}\hat{\Lambda}\right)+
\frac{i}{2}N^{3/2}\Str \hat{J}_k\left(G_2
\hat{T}^{-1}\hat{\Lambda}\hat{T}\hat{\Lambda}+iG_1\hat{\Lambda}
\right)\right\}
\end{eqnarray}
whith $d\mu(T)$ being the invariant (Haar's) measure on the 
graded coset space.
Here we assumed an infinitesemal negative imaginary part to be
included in $s$. Remembering that $\hat{J}_k=J_1\hat{k}
\frac{1+\hat{\Lambda}}{2}+J_2\hat{k}
\frac{1-\hat{\Lambda}}{2}$ 
and making the correspondence:
\begin{equation}\label{corr}
G_1(r)\longrightarrow
\frac{1}{N}\sum_i\frac{\lambda-v_it}{(\lambda-v_it)^2+q^2v_i^2}
\quad \mbox{and} \quad 
G_2(r)\longrightarrow
\frac{1}{N}\sum_i\frac{q v_i}{(\lambda-v_it)^2+q^2v_i^2}
\end{equation}
we see that the obtained expression literally
coincides \cite{noteN} with the averaged generation function derived in the
previous section, that means with 
the generating function Eq.(\ref{genave}) after restricting the integration 
to the saddle-point manifold Eq.(\ref{spm}) and expanding 
with respect to source terms.
Since that generating function
 underlied the expressions Eqs.(\ref{con}) for the two-point spectral
correlation function 
as well as the  expression Eq.(\ref{meand}) for the mean spectral density,
we immediately extract those quantities for our actual problem.
As we already mentioned above, the mean resonance density
is given by $\rho(r)=\frac{1}{\pi}G_2(r)$, and the functional form of the
spectral correlation function is given by the same 
Wigner-Dyson expression
as before, see Eq.(\ref{Dyson}), provided one uses $\rho(r)$ for 
the actual mean resonance density.

Finally, it is necessary to mention that using the same method it is
straightforward to demonstrate that all 
higher spectral  correlation functions
will be given by GOE expressions as well.

\section{Conclusion}
In the present paper we introduced the full-connectivity model of 
a disordered reactance $LC$ network and found that its 
spectral properties in 
the thermodynamic limit $N\to \infty$ could be efficiently investigated
in the framework of a version of Efetov's supersymmetry method
\cite{Efbook,VWZ,my} exploiting a generalized Hubbard-Stratonovich
transformation introduced by us earlier\cite{MF}.
 We also studied spectral properties of regular
pencils of random matrices with i.i.d. entries. In all cases we were
able to derive the mean spectral density as well as to characterize
fluctuation properties of spectra. The models studied turned out to be
a faithful representatives of the Wigner-Dyson universality class.

Actually, we hope that the present paper may provide a convenient
background for the regular investigation of fluctuations in
electric  properties of disordered
$LC$ (more generally, $RL-C$) 
networks, such as the conductance $Y_{AB}$ defined in Eq.(\ref{YY}),
 on-site potentials $V_i$, etc. Actually, 
it is possible to consider
an $LC-$ model of a``banded'' type (i.e. of a large, but finite range, 
see \cite{band} and \cite{FMS} )  and derive \cite{myprep} 
the corresponding d-dimensional nonlinear $\sigma-$model which should 
adequately describe properties of the realistic $LC-$ networks\cite{BBS,Luck3},
including the effects of Anderson localization\cite{loc} 
These issues will be addressed  in forthcoming publications \cite{myprep}.

{\bf Acknowledgements}
The author acknowledges J.M.Luck and E.F.Shender for attracting
his attention to properties of disordered reactance networks
and is grateful to B.A.Khoruzhenko and especially to L.A.Pastur
for their comments concerning pencils of random matrices. 
The work was completed during the author's stay at Max-Planck
Institute for Complex Systems in Dresden as a participant of
the Workshop on "Dynamics of Complex Systems"
whose kind hospitality and generous support is gratefully acknowledged.

The work was supported in part by SFB-237 "Disorder and Large
Fluctuations" as well as by the grant INTAS 97-1342: "Magnetotransport,
localization, interactions and chaotic scattering in low-dimensional
electron systems". 

\appendix
\section{A digression on functional Hubbard-Stratonovich transform,
Eq.(\ref{FHTS})}
In this  Appendix we present a heuristic demonstration of 
the validity of the relation Eq.(\ref{FHTS}) for some kernels ${\cal K}$.
We are not pretending that we are able to specify
under what precise conditions 
our formal manipulations to be true: this interesting 
issue deserves to be
studied seriously and goes much beyond our modest goals.
Instead, our strategy will be a pragmatic one:
to illuminate the origin of relations of that kind. 

Actually, the fact that $\Phi$
is a supervector rather than a usual vector is of little importance for
our consideration, so one may think about it as a usual vector
(or even scalar). The way to incorporate anticommuting variables
in the consideration sketched below 
is described in the Appendix A of \cite{FMS}.

Let us deal then with a real symmetric integral kernel
${\cal K}(\Phi_a,\Phi_b)={\cal K}(\Phi_b,\Phi_a)$. Then, generally 
we expect to exist a set of 
real eigenvalues $\lambda_{\nu}$ and orthogonal and
normalizable set of corresponding real-valued
 eigenfunctions $e_{\nu}(\Phi)$:
\begin{equation}\label{eig}
\int d\Phi_b {\cal K}(\Phi_a,\Phi_b)e_{\nu}(\Phi_b)
=\lambda_{\nu}e_{\nu}(\Phi_a)\quad,\quad \int d\Phi e_{\nu}(\Phi)
e_{\mu}(\Phi)=\delta_{\mu\nu}   
\end{equation}
such that the kernel ${\cal K}$ allows the following representation:
\begin{equation}\label{kereig}
{\cal K}(\Phi_a,\Phi_b)=\sum_{\nu}\lambda_{\nu}e_{\nu}(\Phi_a)
e_{\nu}(\Phi_b)   
\end{equation}
where the summation goes over all $\nu$ such that $\lambda_{\nu}\ne 0$.
Then we can write a formal chain of transformations:
\begin{eqnarray}\label{chain}\nonumber
I&=&\exp\left\{-\frac{1}{2}\sum_{i,j=1}^N{\cal K}(\Phi_i,\Phi_j)\right\}
=\prod_{i,j}\exp\left\{-\frac{1}{2}\sum_{\nu}\lambda_{\nu}e_{\nu}(\Phi_i)
e_{\nu}(\Phi_j)\right\}=\prod_{\nu}\exp\left\{-\frac{\lambda_{\nu}}{2}
\left(\sum_{i=1}^Ne_{\nu}(\Phi_i)\right)^2\right\}\\
&=&\prod_{\nu}\int_{-\infty}^{\infty}\frac{dz_{\nu}}{\sqrt{2\pi}}
\exp\left\{-\frac{z_{\nu}^2}{2}-
i\lambda_{\nu}^{1/2}z_{\nu}\sum_{i=1}^{N}e_{\nu}(\Phi_i)\right\}\\
&=&\int...\int\left(\prod_{\nu}\frac{dz_{\nu}}{\sqrt{2\pi}}\right)
\exp{\left[-\frac{1}{2}\sum_{\nu}z_{\nu}^2-i\sum_{i=1}^N
\left(\sum_{\nu}\lambda_{\nu}^{1/2}z_{\nu}e_{\nu}(\Phi_i)\right)\right]}
\label{last}
\end{eqnarray}
Again, we do not specify conditions when it is allowed to
interchange the summations, products and integrations but just assume
simple-mindedly that the sequence of transformations presented above
 does not lead to divergent expressions.

A form of the expression Eq.(\ref{last}) suggests to consider it
as a definition of a functional integral going over the space of
functions:
\begin{equation}\label{space}
g(\Phi)=\sum_{\nu}\lambda_{\nu}^{1/2}z_{\nu}e_{\nu}(\Phi)\quad,\quad
{\cal D}(g)\equiv \prod_{\nu}\frac{dz_{\nu}}{\sqrt{2\pi}}
\end{equation}
Finally, we introduce the kernel ${\cal C}(\Phi_a,\Phi_b)$ as:
\begin{equation}\label{calc}
{\cal C}(\Phi_a,\Phi_b)=\sum_{\mu}\lambda_{\mu}^{-1}e_{\mu}(\Phi_a)
e_{\mu}(\Phi_b)
\end{equation}
which obviously satisfies:
\begin{equation}\label{del}
\int d\Phi{\cal C}(\Phi_a,\Phi){\cal K}(\Phi,\Phi_b)=\sum_{\mu}
e_{\mu}(\Phi_a)e_{\mu}(\Phi_b)\equiv \delta_g(\Phi_a,\Phi_b)
\end{equation}
where the $\delta_g(\Phi_a,\Phi_b)$ plays the role of a $\delta-$function
in the space spanned by functions $g(\Phi)$, Eq.(\ref{space}):
\[
\int d\Phi_ag(\Phi_a)\delta_g(\Phi_a,\Phi_b)=g(\Phi_b)
\]
Moreover, using Eq.(\ref{space})
it is straightforward to verify that:
\[
\int d\Phi_bd\Phi_ag(\Phi_a){\cal C}(\Phi_a,\Phi_b)g(\Phi_b)=z_{\nu}^2
\]
which finally allows to rewrite Eqs.(\ref{chain},\ref{last})
as:
\begin{equation}\label{FHSTL}
\exp\left\{-\frac{1}{2}\sum_{i,j=1}^N{\cal K}(\Phi_i,\Phi_j)\right\}
=\int{\cal D}(g)\exp\left\{-\frac{1}{2}
\int d\Phi_bd\Phi_ag(\Phi_a){\cal C}(\Phi_a,\Phi_b)g(\Phi_b)-i
\sum_{i=1}^Ng(\Phi_i)\right\}
\end{equation}
which can be brought to the form Eq.(\ref{FHTS}) by a trivial scaling
transformation.

Let us finally note, that our actual kernel Eq.(\ref{kern}) is of a 
separable nature and, moreover, has only few nonzero eigenvalues.
Evidently, in that case the formal 
manipulations Eq.(\ref{chain}) are expected to be most harmless.


\begin{thebibliography}{99}
\bibitem{noteN} An extra factor $N^{1/2}$ appearing in the exponent 
Eq.(\ref{genaverT}) as compared to similar equations of the previous
paragraph is due to our rescaling: $\tilde{\lambda}\to r/N^{1/2}$
which requires also the spectral density to be redefined
appropriately: $\rho(r)=N^{-1/2}\rho(\tilde{\lambda})$
\end{thebibliography}
\end{document}